# Pressure-induced quantum phase transition in the itinerant ferromagnet UCoGa

M.Míšek[1], J.Prokleška[2], P. Opletal[2], P. Proschek[2], J. Kaštil[1], J. Kamarád[1] and V. Sechovský[2]

[1]*Institute of Physics, Academy of Sciences of Czech Republic, v.v.i, Na Slovance 2, 182 21, Prague 8, Czech Republic*

[2]*Faculty of Mathematics and Physics, Charles University, Department of Condensed Matter Physics, Ke Karlovu 5, 121 16, Praha 2, Czech Republic*

In this paper, we report the results of a high pressure study of the itinerant 5f-electron ferromagnet UCoGa. The work is focused on probing the expected ferromagnet-to-paramagnet quantum phase transition induced by high pressure and on the general features of the P-T(-H) phase diagram. Diamond anvil cells were employed to measure the magnetization and electrical resistivity under pressures up to ~ 10 GPa.

At ambient pressure, UCoGa exhibits collinear ferromagnetic ordering of uranium magnetic moments $\mu_U$ ~ 0.74 $\mu_B$ (at 2 K) aligned along the c-axis of the hexagonal crystal structure below Curie temperature $T_C$ = 48K. With the application of pressure, gradual decrease of both, $T_C$ and the saturated magnetic moment, has been observed up to pressures ~ 6 GPa. This is followed by a sharp drop of magnetic moment and a sudden disappearance of the magnetic order at the pressure of 6.5 GPa, suggesting a first-order phase transition, as expected for a clean system. The low temperature power law dependence of the electrical resistivity shows distinct anomalies around the ~ 6 GPa, consistent with the pressure evolution of the magnetic moment and the ordering temperature. The tricritical point of the UCoGa phase diagram is located at approximately ~ 30K and ~ 6GPa.

## I. INTRODUCTION

UCoGa belongs to the group of U*TX* (*T* - transition metal, *X* - p-element) compounds, crystallizing in the hexagonal ZrNiAl - type structure (P-6m2). These uranium intermetallic compounds form for the late transition metals (*T*) and the III-V group p-elements (*X*) [1]. Magnetism in these intermetallics is governed by two key mechanisms – overlap of 5f wave functions of neighboring uranium atoms and hybridization of the uranium 5f states with the valence states of *T* and *X* ligands. Several compounds of this group order ferromagnetically, providing a playground for investigation of the critical behavior of itinerant 5f-electron ferromagnets. The very strong uniaxial magnetic anisotropy locks the U magnetic moments aligned along the c-axis of the hexagonal structure, whereas only Pauli paramagnetism is observed in the perpendicular directions.

In recent years, the critical behavior of itinerant electron ferromagnets has enjoyed a renewed interest sparked by the observation of many unusual phenomena (anomalous non-Fermi liquid behavior, magnetically mediated Cooper pairing, etc.) in the vicinity of loss of the long-range ferromagnetic order[2]. Contrary to the case of antiferromagnetism, where novel states emerge at a quantum critical point (QCP, see e.g. [3,4]), the ferromagnet-to-paramagnet quantum phase transitions (QPT) have been found to be the cause of these phenomena, opening an entirely new chapter of investigation of ferromagnetic materials ranging from the classical and quantum description of low temperature order-disorder transitions to fundamental question of existence of ferromagnetic QCP.

Particularly, the pressure induced ferromagnet-to-paramagnet transition at low temperatures has been discussed within different scenarios, where quantum phase transitions may take place. In clean systems at low temperatures, two scenarios are possible: A first-order ferromagnet-to-paramagnet transition, or the appearance of an inhomogeneous magnetic phase between the ferromagnetic and paramagnetic state[2]. There are only few known U-based ferromagnets displaying the change of the order of transition, demonstrating the appearance of the first-order transition by driving the system close to the loss of the long-range magnetic order[2].
In this paper, we present the observation of tricritical point and possible presence of QPT in UCoGa, extending the available (and limited) knowledge of systems showing discontinuous transition in the vicinity of the pressure induced ferromagnet-to-paramagnet transition.

## II. EXPERIMENTAL

The UCoGa single crystals were pulled from a stoichiometric melt by Czochralski method in a tri-arc furnace with a rate of 12 mm/h. The grown crystals were wrapped in a tantalum foil and annealed in vacuum ($10^{-6}$ mbar) inside quartz tube for 7 days in 900°C. The structure and composition of the single crystals were checked by X-ray Laue method (Laue diffractometer made by Photonic Science), X-ray powder diffraction (Bruker AXS D8 Advance X-ray difractometer with Cu X- ray tube) and energy dispersive X-ray spectroscopy (scanning electron microscope Tescan Mira I LMH). Samples for specific experiments were cut by wire saw from the single crystal oriented with Laue diffractometer. Magnetic measurements were performed using a MPMS 7 XL SQUID magnetometer (Quantum Design) with the magnetic field applied along the c-axis. For the application of pressure, the "turnbuckle"- type diamond anvil cell[5] was used. Transport properties were measured in a PPMS (Quantum Design) apparatus with an in-house designed Merryll-Basset type diamond anvil cell. In both cases, the Ruby pressure scale and Daphne 7474 pressure transmitting medium were used.

## III. RESULTS

For the high pressure measurements we have selected small pieces of larger crystal with residual resistivity $\rho_0 \approx 10\mu\Omega\cdot$cm, which can be classified as clean within the given family of compounds [2]. At ambient pressure, the temperature dependence of electrical resistivity shows clear cusp-like anomaly in the vicinity of $T_C$ and power-law scaling at low temperatures (see Fig 1a).

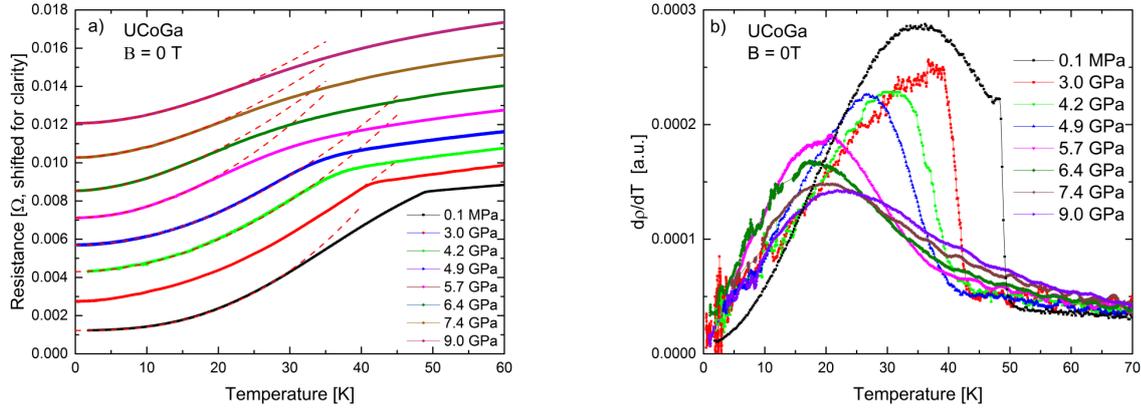

**Fig.1. a)** Temperature dependence of the electrical resistivity under applied pressures. Individual curves are shifted for clarity, dashed lines indicate power law fit for each pressure (see text for details). **b)** Temperature dependence of derivative of the electrical resistivity under applied pressures.

With applied pressure, clear decrease of the $T_C$ is visible, accompanied by smearing out of the $T_C$-related anomaly at higher pressures. This behavior is better seen in Fig. 1b), where the temperature derivative of the resistivity is shown. The step-like change (at ambient and low pressures) in the derivative is turned into a broad bump at 5-6 GPa which remains unchanged at higher pressures. For better understanding of pressure evolution, the power law fit ($\rho = \rho_0 + AT^n$) at low temperatures (see Fig. 1a) is used for description. At low pressures the resistivity exponent is close to 2, in agreement with the ferromagnetic ground state of the compound. With increasing pressure, the abrupt change of the evolution in the vicinity of 6 GPa indicates the order-disorder transition (see Fig. 2a). The resistivity exponent drops close to 5/3 at the presumed quantum phase transition as expected by the theory of 3D spin fluctuations [6]. Similar values were observed in the isostructural URhAl [7] and UCoAl [8] compounds in the vicinity of the suppression of long-range ferromagnetism, indicating the importance of spin fluctuations in the magnetism of this family of compounds.

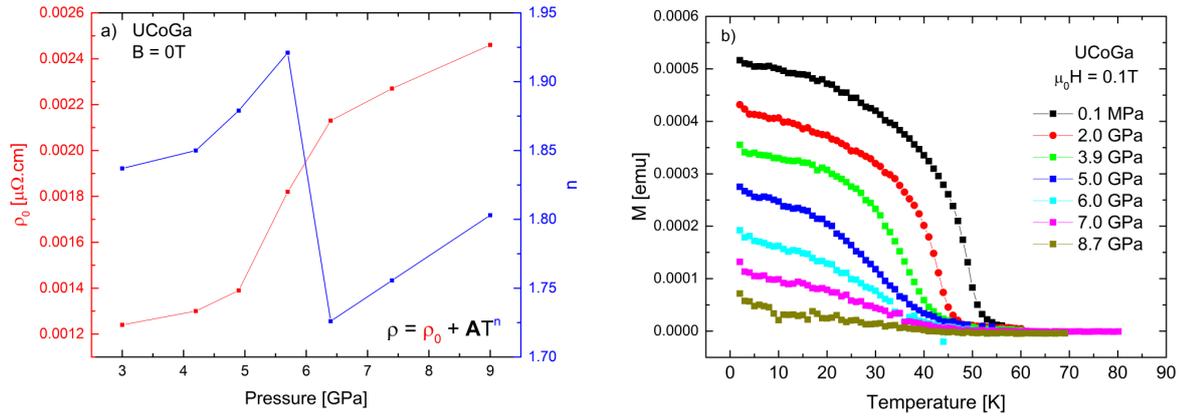

**Fig.2. a)** Pressure dependence of coefficients in the power law – n (blue) and residual resistivity $\rho_0$ (red). **b)** The temperature dependence of magnetization as measured in 0.1T field applied along the c-axis for various applied pressures.

This is in accordance with the increase of residual resistivity at higher pressures in non-magnetic state due to the additional scattering on disordered magnetic moments.

To confirm the sudden suppression of magnetism and loss of long-range ordering, the pressure evolution of magnetic properties was measured directly. The temperature dependence of the magnetization measured at 0.1T (field cooled) demonstrates well defined $T_C$ value at low pressures (see Fig. 2b), being suppressed to ~30 K at ~6 GPa. At higher pressures there is only very weak, almost paramagnetic-like increase of magnetization, indicating the loss of long-range magnetic order. The investigation of saturated magnetization measured at 0.5 T (i.e. well above coercive field of UCoGa) shows continuous decrease with increasing temperature (for given pressure), as expected for an itinerant ferromagnet (see Fig. 3a). The saturated magnetization at 2K shows continuous suppression to about 2/3 of ambient value, followed by a rapid drop in the vicinity of critical pressure.

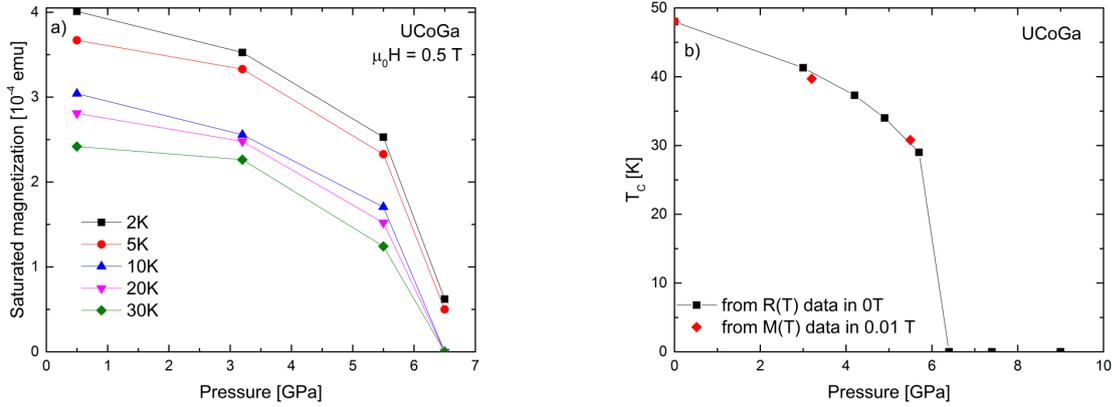

**Fig.3. a)** Pressure variation of the saturated magnetization at various temperatures in 0.5 T field applied along the c-axis. **b)** Pressure dependence of the ordering temperature determined from the measurements of temperature dependence of magnetization and electrical resistivity.

Due to the significant anisotropy of crystal structure, a general note to the application of hydrostatic pressure should be mentioned. As discussed in the Introduction, UCoGa has the layered ZrNiAl-type structure, being strongly anisotropic, both in magnetic and mechanical properties. The directions within the basal plane are soft (the compressibility is about 3-times softer than along the c-axis) and the shortest U-U distance is within the sheets as well, leading to a synergy effect under applied hydrostatic pressure – the closer the uranium atoms are, the larger the effect of the pressure is (for details see e.g. [9] and references therein). This is the prevailing effect of hydrostatic pressure and for moderate U-U distances (UCoGa case) it does not change within the range of pressure used in this study, consequently this allows the use of the correlation between the pressure, tuning of electronic structure and control parameter.

## IV. CONCLUSIONS

The obtained results can be summarized in the proposed magnetic phase diagram as shown in Fig. 3b). Both, transport and magnetic measurements reveal continuous suppression of transition temperature down to 30 K in 6 GPa, which we identify as the tricritical point (TCP), followed by rapid change of observed quantities (both on transport and magnetic properties) related to the loss of long-range magnetic ordering. The phase diagram follows the expectations for the clean itinerant ferromagnet (see e.g. [2]), i.e. continuous suppression of ordering with increasing pressure down to TCP, where rapid drop in observables is expected in accordance to the presence of first-order transition. Similar behavior was observed in URhAl [7] and UCoAl [10]. The presented experimental data are in agreement with comparable literature data and theoretical predictions[2]. Temperature of the TCP is rather high as compared to similar compound URhAl, nevertheless, the sample used in study [7] has significantly higher residual resistivity (65 μΩ·cm) and the URhAl has lower magnetic moment. From theory it is known that both, increase of disorder (residual resistivity) and smaller magnetic moment, affect negatively the tricritical temperature [11]. It should be expressed that this study, together with the previously reported works [7, 10] indicate, that the observed phenomenon may be common in the U$TX$ compounds with ZrNiAl-type structure. These compounds with easily scalable Curie temperatures and magnetic moments can be used as a playground for the investigation of quantum critical phenomena and verification of related scaling properties. In order to unambiguously confirm the proposed scenario, a detailed investigation of the vicinity of TCP (i.e. search for field induced 'wings') is desirable.

**ACKNOWLEDGMENTS**

This work is part of the research program GACR 16-06422S which is financed by the Czech Science Foundation. Experiments were performed in the Magnetism and Low Temperature Laboratories, which is supported within the program of Czech Research Infrastructures, project no. LM2011025.